# Topological-Charge-Enabled Photonic Doping in ENZ Media

Zhicheng Xiong[1], Fei Sun[1,*], Weiqi Yuan[1], Yichao Liu[1], Shuai Zhang[2], Zhihui Chen[1], and Mingda Zhang[1]

[1]*Key Lab of Advanced Transducers and Intelligent Control System, Ministry of Education and Shanxi Province, College of Physics and Optoelectronics, Taiyuan University of Technology, Taiyuan, 030024 China*

[2] *Department of Electronic Systems, Aalborg University, 9220 Aalborg, Denmark*

[*]*sunfei@tyut.edu.cn*

**ABSTRACT**

Conventional photonic doping schemes predominantly employ circular or rectangular dielectric dopants with zero topological charge, where the effective permeability can only be tuned through material selection and geometric scaling, resulting in limited design flexibility. In this work, topological structures are introduced into dielectric dopants by embedding internal holes to generate nonzero topological charge. Based on this concept, a theoretical model is established to describe the effective permeability of photonic doping systems with nonzero topological charge, and the underlying mechanisms governing topological-charge-dependent transmission are systematically elucidated. The results demonstrate that engineering nonzero topological charge through the number, shape, size and position of internal holes within dielectric dopants enables flexible manipulation of the internal magnetic field distributions, thereby providing precisely control over the effective permeability, as well as the resonance frequency and spectral linewidth of the transmission spectrum. The proposed multi-dimensional photonic doping strategy, integrating topological-charge engineering with geometric design, substantially enriches the available degrees of freedom for dispersion engineering and provides a versatile platform for advanced functional photonic devices.

## 1. Introduction

In recent years, electromagnetic metamaterials [1–6] have undergone rapid development, breaking the inherent restrictions on electromagnetic parameters found in conventional natural materials and enabling precise manipulation of permittivity and permeability. Among them, epsilon-near-zero (ENZ) media have become a research frontier and hotspot in the field of electromagnetic and optical wave manipulation due to their unique properties such as infinite wavelength, zero phase delay, field confinement, and space-time decoupling [7–12]. Relying on these extraordinary effects, ENZ materials can realize numerous exotic physical phenomena and functional applications, including arbitrarily shaped flexible waveguides [13,14], enhanced nonlinear responses [15], enhanced Purcell factors [16], coherent perfect absorption [17,18], novel optical resonant cavities [19], ultra-thin invisibility devices [20], ultra-thin absorbers [21], optical funnels [22], and omnidirectional scattering suppression [23]. However, the equivalent electromagnetic parameters of ENZ media are fixed once fabricated, lacking tunability, which limits the applications of ENZ media in reconfigurable electromagnetic devices [24]. To address this issue, the concept of photonic doping offers an effective approach to tailor the electromagnetic parameters of ENZ media [25]. By doping with one or more macroscopic dopants in an ENZ host, photonic doping endows the whole photonic doping systems with an overall effective permittivity $\varepsilon_{\text{eff}} \approx 0$ and enables continuous tuning of the effective permeability ($\mu_{\text{eff}}$) from negative values, through zero, to infinity. Moreover, the equivalent electromagnetic parameters of the photonic doping systems are independent of the position and orientation of the dopants. Photonic doping expands the application paradigm of the traditional effective medium theory and overcomes the constraint that the sizes of dopants must be much smaller than the operating wavelength, allowing dopants to be comparable to or even larger than the operating wavelength (i.e., macroscopic doping). Accordingly, the equivalent electromagnetic parameters of the photonic doping systems can be tailored by adjusting the shape, material, and size of macroscopic dopants, providing a new route for the design of advanced photonic devices.

Based on photonic doping, incorporating different dopants into an ENZ host enables continuous tuning of the effective permeability $\mu_{\text{eff}}$ from zero to infinity for the overall structure (i.e., regulating the equivalent impedance of the photonic doping systems), thereby precisely controlling the transmission spectrum of electromagnetic waves in ENZ media. For instance, doping an ENZ host with dielectric dopants featuring gain or loss gives rise to non-Hermitian doping [26], while doping with Kerr dielectric inclusions provides an additional tuning degree of freedom to dynamically modulate the transmission spectrum via the intensity of incident light [27]. These approaches offer additional controllable degrees of freedom for transmittance engineering. Furthermore, doping with rectangular dielectric impurities of varying sizes can generate a "coded" transmission spectrum, namely dispersion coding [28]. In addition, photonic doping has motivated a range of advanced research topics, including bound states in the continuum independent of geometric symmetry [29], geometry-independent antennas with radiation frequencies [30], metamaterial processors capable of performing calculus operations on transmitted electromagnetic waves [31], highly uniform electric field enhancement at deep-subwavelength scales [32], and high-performance impedance matching layers for arbitrary interfaces [33]. However, existing studies are restricted to dielectric dopants with zero topological charge, such as circular and rectangular dopants. As a rapidly emerging degree of freedom for electromagnetic wave control in recent years, topological charge has exhibited powerful capabilities in tailoring vortex beams, vector optical fields, and light field manipulation [34]. Despite its great potential in wavefront engineering and photonic design, the critical role of topological charge in regulating the overall transmittance characteristics of the photonic doping systems remains largely unexplored, which limits the flexibility of photonic doping in transmission spectrum manipulation.

Conventional transmission spectrum modulation strategies by photonic doping can only tune the peak and valley features of the transmission spectrum by changing the materials and geometric dimensions of dopants, which leads to a single modulation dimension and strong dependence on material and profile parameters. By introducing topological charge as a novel tuning degree of freedom, different topological charge states can be achieved simply by creating different numbers of holes inside the dopant while keeping the material type and outer contour size of the dielectric dopant completely unchanged. The transmission spectrum characteristics of the photonic doping systems can then be flexibly tuned by the number and size of the holes. This approach breaks the inherent dependence of traditional modulation on material type and overall dimensions, effectively expands the modulation scope of photonic doping, and significantly improves the design freedom and tuning flexibility of dispersion coding. Accordingly, this study proposes photonic doping in ENZ media using square dielectric dopants with nonzero topological charge, which introduces topological charge as a new tuning dimension into the photonic doping system and systematically investigates the influence of nonzero topological charge on the effective electromagnetic parameters and transmission spectrum characteristics of the photonic doping systems. Specifically, this study first establishes a theoretical model for calculating $\mu_{\text{eff}}$ in photonic doping systems with nonzero topological charge and explores the tuning law of topological charge $g$ on the transmission spectrum. Then, focusing on the typical case of topological charge $g = 1$, this study deeply analyzes the tuning laws and physical mechanisms of the size, position, and shape of internal holes in square dielectric dopants on the frequency shift and transmission spectrum broadening. These results reveal the intrinsic tuning mechanism of topological charge in photonic doping systems and further expand the modulation dimensions and application potential of photonic doping technology.

## 2. Photonic Doping Theory for Dopants with Nonzero Topological Charge

To reveal the modulation mechanism of topological charge upon photonic doping characteristics, this work first establishes a theoretical model describing the effective permeability of photonic doping systems carrying nonzero topological charge. As illustrated in the left panel of Fig. 1(a), we investigate a two-dimensional (2D) doped ENZ body with area $A$, embedded with a 2D dielectric dopant characterized by an area $A_{\mathrm{die}}$, relative permittivity $\varepsilon_{\mathrm{d}}$, relative permeability $\mu_{\mathrm{d}}$, and topological charge $g$. The topological charge value is determined by the number of inner holes: $g$=0 corresponds to a hole-free dopant, $g$=1 for a single hole, $g$=2 for two holes, and so on. The total area of the internal ENZ holes is $A_{\mathrm{hole}}$, and the area of the total doped region $A_{\mathrm{dopant}}$ follows the relation: $A_{\mathrm{die}}+A_{\mathrm{hole}}=A_{\mathrm{dopant}}$. The electromagnetic fields obey the curl equations derived from the differential form of Maxwell's equations:

$$\nabla \times \boldsymbol{H} = -i\omega\varepsilon\boldsymbol{E}. \tag{1}$$

In the ENZ host and holes filled with the ENZ medium, the permittivity $\varepsilon$ approaches zero, so the curl of the magnetic field is also zero. When TM-polarized electromagnetic waves are incident on the 2D structure, the magnetic field has only the $z$-component:

$$\boldsymbol{H} = H(x,y)\hat{z}, \tag{2}$$

substituting (2) into (1) yields:

$$\nabla \times (H\hat{z}) \approx 0, \tag{3}$$

after simplification, we obtain:

$$\nabla H = 0. \tag{4}$$

The above derivation further verifies that the TM-polarized magnetic field is uniformly distributed in the ENZ host and the holes filled with the ENZ medium [16]. Next, Faraday's law of induction is applied to the total doped ENZ body:

$$\begin{cases} \oint_{\partial A} \boldsymbol{E}\cdot \mathrm{d}\boldsymbol{l} = \mathrm{i}\omega \iint_A \boldsymbol{B}\cdot \mathrm{d}s = \mathrm{i}\omega\mu_0 H_0 \left( A_{\mathrm{ENZ}} + \eta + \mu_{\mathrm{d}} \iint_{A_{\mathrm{die}}} \psi^{\mathrm{d}} \mathrm{d}s \right) \\ \eta = \begin{cases} 0, g=0 \\ \sum_{n=1}^{g} q_n A_{hole,n}, g \geq 1 \end{cases} \end{cases}, \tag{5}$$

where $A_{\mathrm{ENZ}}$ represents the area of the ENZ host medium surrounding the dielectric dopant, and $\eta$ quantifies the contribution of ENZ holes to the equivalent magnetic permeability; $A_{hole,\,n}$ is the area of each hole filled with the ENZ medium inside the dielectric dopant, while $q_n$ is defined as the ratio of magnetic field intensity inside the ENZ holes to that in the outer ENZ host. $\mu_{\mathrm{d}}$ is the relative permeability of the dielectric dopant, and $\psi^{\mathrm{d}}$ is the normalized magnetic field amplitude of the dopant evaluated at the ENZ host boundary.

We introduce a homogeneous equivalent medium model shown in the right panel of Fig. 1(a). This equivalent domain shares an identical outer contour with the original 2D ENZ host and possesses an effective permeability $\mu_{\mathrm{eff}}$, such that it produces identical scattering responses as the doped 2D ENZ body illustrated in the left panel of Fig. 1(a) (i.e., the scattered electromagnetic fields match exactly under identical incident waves). In accordance with classical photonic doping theory, the equivalent permittivity of this homogenized medium still satisfies $\varepsilon_{\mathrm{eff}}\approx 0$ [26]. For a dopant with topological charge $g$, we derive the overall effective permeability by enforcing magnetic field continuity at the structural boundary: the magnetic field magnitude within the ENZ host regions of both the doped ENZ body and the homogenized equivalent medium equals $H_0$. Under this constraint, Faraday's law of induction takes the form:

$$\oint_{\partial A} \boldsymbol{E} \cdot \mathrm{d}\boldsymbol{l} = i\omega\mu_0\mu_{\mathrm{eff}} H_0 A. \tag{6}$$

By comparing Eq. (5) and Eq. (6), the expression for the effective permeability of the photonic doping system with nonzero topological charge is obtained:

$$\begin{cases} \mu_{\mathrm{eff}} = \left( A_{\mathrm{ENZ}} + \eta + \mu_{\mathrm{d}} \iint_{A_{\mathrm{die}}} \psi^{\mathrm{d}} \mathrm{d}s \right) / A \\ \eta = \begin{cases} 0, g = 0 \\ \sum_{n=1}^{g} q_n A_{hole,n}, g \geq 1 \end{cases} \end{cases} . \tag{7}$$

Compared with the zero-topological-charge system with $g$=0 (the $g$=0 case in Eq. (7)), the effective permeability formula derived for systems with nonzero topological charge ($g \neq 0$) contains an extra contribution term $\eta$ originating from the ENZ holes. As indicated by Eq. (7), nonzero topological charge $g \neq 0$ serves as an extra degree of freedom to modulate the effective permeability. The introduced topological charge $g$ collaboratively tunes the effective permeability of the homogenized photonic doping systems via the coupled interplay of four factors: the area of each individual ENZ hole $A_{hole,g}$, topological charge $g$, the area of the ENZ host $A_{\mathrm{ENZ}}$, and the internal magnetic field distribution $\psi^{\mathrm{d}}$ within the dielectric dopant. Subsequent numerical simulations will further uncover the impact of these coupled factors and systematically explore how diverse geometric parameters govern the effective permeability.

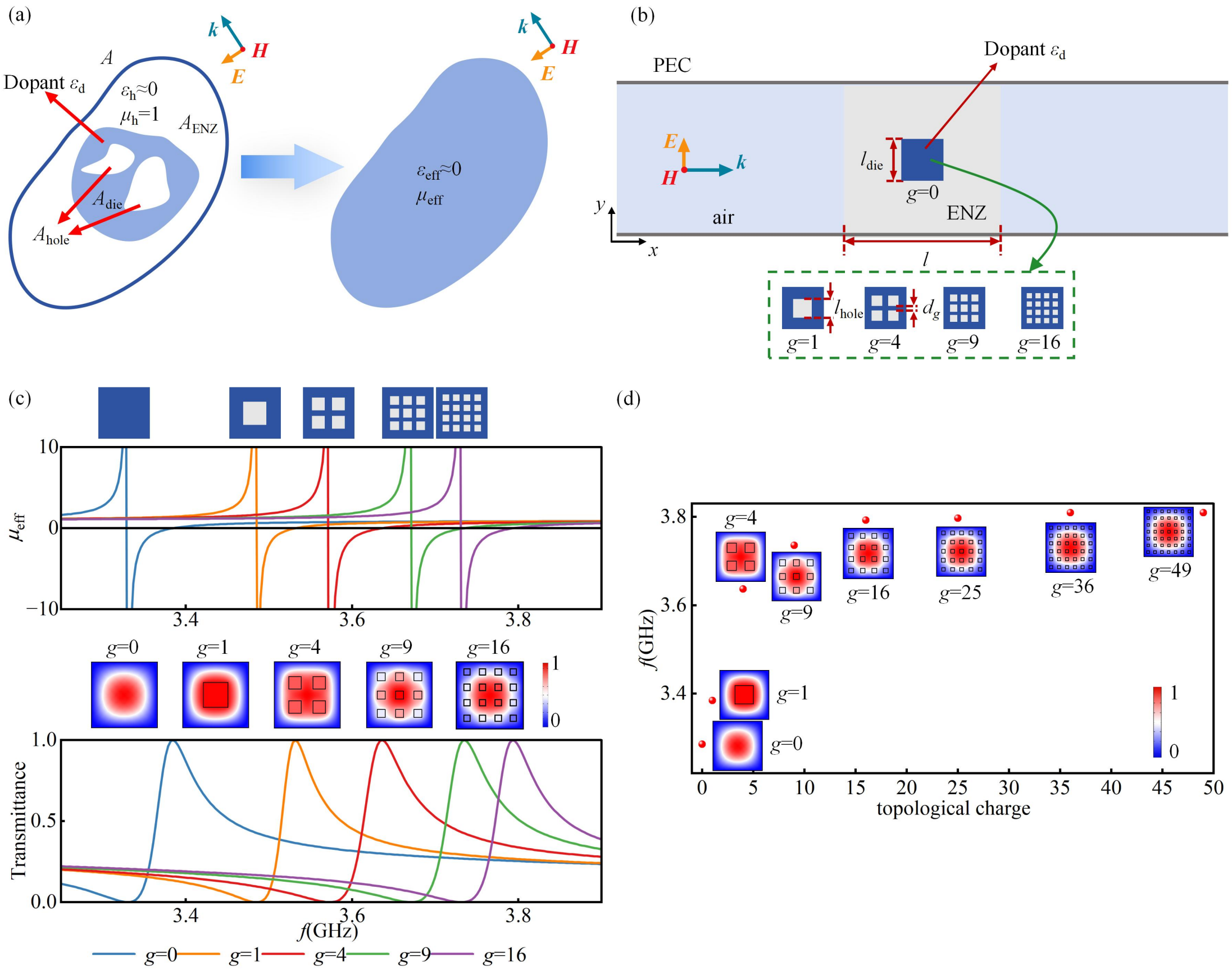

**Fig. 1.** (a) Schematic diagram for calculating the effective electromagnetic parameters of a 2D ENZ

medium doped with impurities of nonzero topological charge $g\neq 0$: The left panel shows the actual physical structure, where a dielectric dopant with topological charge $g$=2 and relative permittivity $\varepsilon_d$ is embedded in a 2D ENZ host with relative permittivity $\varepsilon_h\approx 0$ and relative permeability $\mu_h$=1. The right panel shows the theoretically calculated homogeneous equivalent medium structure (with the same outer boundary as the 2D ENZ host), which has an effective permittivity $\varepsilon_{eff}\approx 0$ and effective permeability $\mu_{eff}$. When TM-polarized electromagnetic waves are incident on the two structures at any angle, the scattered fields produced by them are identical. (b) A parallel-plate waveguide operating in the TM fundamental mode (TEM mode), where the electric field is polarized along the $y$-direction and the magnetic field along the $z$-direction. A square ENZ substrate with side length $l$ ($\varepsilon_h\approx 0$, $\mu_h$=1) is placed in the middle of the waveguide. A nonmagnetic square dielectric dopant with area $A_{die}=l_{die}\times l_{die}$ and relative permittivity $\varepsilon_d$ is embedded in the substrate (where $g$=0). On this basis, different topological charge $g$ dopants can be obtained by embedding one or more square holes (uniformly distributed with equal area, total area $A_{hole}=l_{hole}\times l_{hole}$) inside the square dielectric dopant and filling the holes with the same ENZ medium as the background (the edge spacings $d_4$, $d_9$, and $d_{16}$ correspond to $g$=4, 9, and 16, respectively). (c) Variation curves of the effective permeability (upper) and transmittance (lower) with the incident electromagnetic wave frequency $f$ for different topological charge values of the dopant in the rectangular waveguide structure corresponding to Fig. 1(b), when TM fundamental mode electromagnetic waves are incident from left to right. The upper and lower insets show the schematic diagrams of the dielectric dopant structures with different topological charges and the amplitude distribution of the normalized $z$-component of the magnetic field in the dielectric dopant at the transmittance peak frequency, respectively. In the simulations, $\varepsilon_d$=37, $l=\lambda_p$, $l_{die}=0.108\lambda_p$, $l_{hole}=0.04\lambda_p$, $d_4=0.02\lambda_p$, $d_9=0.017\lambda_p$, $d_{16}=0.015\lambda_p$, and $\lambda_p$=0.1 m. (d) Variation of the transmittance peak frequency with the topological charge $g$ of dopants (the inset shows the amplitude distribution of the normalized $z$-component of the magnetic field inside the dopant at the corresponding frequency). When topological charge $g$=25, 36, and 49, the edge spacings between holes are $d_{25}=0.012\lambda_p$, $d_{36}=0.0103\lambda_p$, and $d_{49}=0.0093\lambda_p$, respectively, and the other parameters are consistent with those in Fig. 1(c).

To streamline numerical investigations focused on transmission spectrum features instead of arbitrary scattering from the general geometry in Fig. 1(a), we perform all subsequent simulations on the 2D parallel-plate waveguide structure depicted in Fig. 1(b). This platform is widely employed to study transmittance responses and dispersion coding functionalities [28, 31, 32]. The waveguide is bounded by perfect electric conductors at its top and bottom walls and operates in the fundamental TM mode. As sketched in the inset of Fig. 1(b), electromagnetic waves propagate along the +$x$ direction, featuring a $y$-polarized electric field and $z$-polarized magnetic field. The waveguide is split into three distinct sections: air fills the input and output channels, whereas the central grey region corresponds to the ENZ host medium characterized by $\varepsilon_h\approx 0$ and $\mu_h$=1. A nonmagnetic square dielectric dopant of side length $l_{die}$ and relative permittivity $\varepsilon_d$ is embedded at the center of the ENZ body. By perforating $g$ identical, uniformly distributed square inner holes of total ENZ hole area $A_{hole}$ within the dopant and filling these holes with the same ENZ host medium, dopant geometries with distinct topological charge values $g$ can be fabricated.

Numerical simulations are performed on the waveguide model illustrated in Fig. 1(b), and the simulated effective permeability and transmittance results are plotted in Fig. 1(c). When the total doped region area $A_{\text{dopant}}$ and total ENZ hole area $A_{\text{hole}}$ are kept constant, increasing the topological charge $g$ synchronously shifts the curves of both effective permeability and transmission spectrum to higher frequencies, where the characteristic resonant points of the two curves exhibit a one-to-one correspondence. Specifically, the structure behaves as a perfect magnetic conductor with vanishing transmittance as the effective permeability approaches infinity. At zero effective permeability, the system forms an epsilon-and-mu-near-zero medium with negligible transmission loss and unity transmittance, which verifies the validity of Eq. (7). As visualized by the embedded magnetic field contour plots in Fig. 1(c), the magnitude of resonant frequency shift is determined by the severity of internal magnetic field distortion within the dielectric dopant compared with the $g$=0 reference case. For $g$=1, the resonant magnetic field distribution bears the closest resemblance to the $g$=0 configuration, resulting in the minimum frequency shift. In comparison, the strongest magnetic field distortion takes place at $g$=16 and yields the largest frequency offset, intuitively demonstrating that resonant frequency shifts originate from the reconstruction of internal magnetic field profiles.

Further analysis of Fig. 1(d) indicates that as the topological charge $g$ increases sequentially to 16, 25, 36, and 49, the overall transmission spectrum shift gradually saturates and exhibits negligible variation upon further raising $g$. The underlying physical mechanism can be explained as follows: when the total doped region area $A_{\text{dopant}}$ and total ENZ hole area $A_{hole}$ are held constant, a larger topological charge $g$ reduces the size of each individual hole. This weakens the holes' perturbation on the internal magnetic field of the dielectric dopant and eventually stabilizes the field profile inside the dopant. Comparing the inset magnetic field contour plots under different topological charge values, the field distribution inside the dielectric dopant barely changes for $g \geqslant 16$. This phenomenon locks the transmission spectrum position and demonstrates the saturation characteristic of topological-charge-mediated modulation.

In summary, topological charge $g$ serves as a pivotal parameter governing the evolution of both effective permeability and transmission spectrum for the waveguide system. When the total doped region area $A_{\text{dopant}}$ and total ENZ hole area $A_{\text{hole}}$ are held constant, raising topological charge $g$ shifts the transmission spectrum to higher frequencies, with the magnitude of frequency shift determined by the degree of internal magnetic field distortion within the dielectric dopant. Specifically, the magnetic field distortion and the resulting resonant frequency shift attain their minimum values at topological charge $g$=1, while both quantities reach maxima at topological charge $g$=16. Tuning topological charge $g$ within the range from 0 to 16 yields remarkable resonant frequency tunability. In contrast, for topological charge $g \geq 16$, resonant frequencies are locked around a constant value with only tiny fluctuations and no longer shift upward owing to the modulation saturation effect; further increases in topological charge thus cannot produce meaningful tuning effects.

## 3. Tuning Laws of Geometric Parameters in Photonic Doping Structures with Topological Charge $g$=1

To overcome the limitations of traditional photonic doping strategies that realize dispersion coding solely by modifying dopant materials and dimensions, and to further enrich the tunable degrees of freedom of photonic doping after uncovering the core modulation capability of topological charge over transmittance responses, this section selects the representative case of topological charge $g$=1 and systematically investigates how the geometric parameters of internal ENZ holes and dielectric dopant area $A_{\mathrm{die}}$ govern transmittance behaviors (Figs. 2 and 3). The theoretical framework and baseline parameter configurations utilized here remain identical to those adopted in Fig. 1(b); only the geometric dimensions of internal ENZ holes and dielectric dopants are modified in the subsequent simulations.

### 3.1 Tuning Laws of Adjustable Total Doped Region Area

Figure 2(a) shows the simulation results obtained by tuning the ENZ hole side length $l_{\mathrm{hole}}$ while keeping the dielectric dopant area $A_{\mathrm{die}}$ constant. Arranged from top to bottom within this subplot are contour plots of the normalized $z$-component magnetic field distribution at each transmittance peak for different values of ENZ hole side length $l_{\mathrm{hole}}$, alongside the corresponding curves of transmittance, effective permeability $\mu_{\mathrm{eff}}$, and magnetic field intensity ratio $q_n$ plotted as functions of incident frequency. As ENZ hole side length $l_{\mathrm{hole}}$ increases, the magnetic energy domains confined within the dielectric dopant are gradually compressed. The resonant peaks initially shift slightly toward lower frequencies alongside mild spectral broadening. As the ENZ hole side length $l_{\mathrm{hole}}$ continues to increase further, the resonant peaks migrate to higher frequencies, accompanied by drastically expanded spectral linewidths. The magnetic field intensity ratio $q_n$ exhibits distinct resonant peaks at the resonant frequencies of $\mu_{\mathrm{eff}}$ and decays to unity far from these resonant points. To intuitively illustrate how the transmittance evolves with ENZ hole side length $l_{\mathrm{hole}}$, Fig. 2(c) provides a two-dimensional color contour map that characterizes transmittance as a joint function of ENZ hole side length $l_{\mathrm{hole}}$ and incident frequency $f$. As ENZ hole side length $l_{\mathrm{hole}}$ grows, the high-transmittance (red zones) initially shift downward to low frequencies before moving upward to higher bands. Spectral broadening remains negligible at small ENZ hole areas, as reflected by the narrow vertical span of red high-transmittance regions. As the ENZ hole side length $l_{\mathrm{hole}}$ further increases, the spectral broadening effect becomes drastically pronounced.

Combined with Eq. (7), the magnetic fluxes confined within the ENZ host, dielectric dopant, and internal ENZ hole regions satisfy a flux balance condition at zero effective permeability $\mu_{\mathrm{eff}}$: $\Phi_{\mathrm{ENZ}}+\Phi_{\mathrm{die}}+\Phi_{\mathrm{hole}}=0$, where $\Phi_{\mathrm{ENZ}}$, $\Phi_{\mathrm{die}}$ and $\Phi_{\mathrm{hole}}$ denote the magnetic fluxes through the ENZ host, dielectric dopant, and internal ENZ holes, respectively. The top contour plots in Fig. 2(a), which display the normalized $z$-component magnetic field distributions inside doped ENZ body at each transmittance peak for different ENZ hole side lengths $l_{\mathrm{hole}}$, reveal the following physical behavior: the magnetic fields localized in the ENZ holes and partial dielectric dopant areas (blue regions) counteract the fields in the remaining dielectric dopant segments and ENZ host (red regions), resulting in a net zero total magnetic flux over the entire doped ENZ body.

When the dielectric dopant area $A_{\mathrm{die}}$ is fixed, increasing the ENZ hole area $A_{\mathrm{hole}}$ expands the total doped region area $A_{\mathrm{dopant}}$ and simultaneously reduces the area of the surrounding ENZ host $A_{\mathrm{ENZ}}$. As the spatial coverage of red regions shrinks, the total magnetic flux that needs to be mutually canceled between the dopant and ENZ holes decreases. Larger ENZ holes produce stronger opposing magnetic flux, alleviating the flux cancellation load borne by the dielectric dopant. To maintain the global flux balance, the zero-crossing points of effective permeability deviate from the resonant frequencies, which weakens the resonant field strength inside the dielectric dopant. The compression and reconstruction of internal magnetic field profiles driven by enlarged hole geometries collectively induce both resonant frequency shifts and spectral broadening of transmittance curves. A critical area $A_c$ divides the response into two separate modulation regimes. If the ENZ hole area $A_{\mathrm{hole}}$ remains smaller than critical area $A_c$, the ENZ holes only impose weak field compression on the dielectric dopant. The dopant still dominates magnetic flux cancellation and sustains intense resonant fields, which leads to minor low-frequency shifts and negligible spectral broadening of transmission spectrum. Once ENZ hole area $A_{\mathrm{hole}}$ surpasses critical area $A_{\mathrm{c}}$, the enhanced compressive effect of holes and the reduced total cancelable flux together transfer the primary flux-canceling role to the ENZ holes. The resonant field amplitude within the dielectric dopant decays sharply, accompanied by significant spectral broadening of transmission spectrum.

Building on the field evolution laws revealed in Fig. 2(a), we further investigate the physical factors that determine the critical area $A_{\mathrm{c}}$. Figs. 2(d) and 2(e) illustrate how the transmittance peak frequency evolves with the ENZ hole side length $l_{\mathrm{hole}}$ under different dielectric dopant areas $A_{\mathrm{die}}$ and relative permittivities $\varepsilon_{\mathrm{d}}$: each curve exhibits an initial decline followed by an upward trend, and the ENZ hole side length $l_{\mathrm{hole}}$ value at each curve's minimum corresponds to the critical area $A_{\mathrm{c}}$ (marked by blue shaded regions). As the dielectric dopant area $A_{\mathrm{die}}$ increases, critical area $A_{\mathrm{c}}$ rises monotonically; in contrast, critical area $A_{\mathrm{c}}$ stays unchanged when adjusting the dopant's relative permittivity $\varepsilon_{\mathrm{d}}$. The underlying physical mechanism, supported by Fig. 2(f), is explained below. A larger dielectric dopant area $A_{\mathrm{die}}$ reduces the overall amount of magnetic flux that needs mutual cancellation and alleviates the spatial compression imposed by internal ENZ holes. Meanwhile, the magnetic flux of the dielectric dopant $\Phi_{\mathrm{die}}$ increases, while the hole flux $\Phi_{\mathrm{hole}}$ decreases at a slower rate as the hole size expands. Consequently, a larger critical threshold $A_{\mathrm{c}}$ is required for $\Phi_{\mathrm{hole}}$ to dominate flux cancellation. At this threshold, the primary flux-canceling role transfers from the dielectric dopant to the ENZ holes, weakening the internal resonant field intensity. On the other hand, tuning $\varepsilon_{\mathrm{d}}$ only shifts transmission spectrum positions without altering the area ratios of each constituent region within the photonic doping structure. For this reason, $A_c$ remains fixed, and only resonant frequency drift can be observed.

Figure 2(b) shows the simulation results obtained with a fixed ENZ hole area $A_{\mathrm{hole}}$ while varying the dielectric dopant area $A_{\mathrm{die}}$. The subplot layout is identical to that of Fig. 2(a). As the dielectric dopant area $A_{\mathrm{die}}$ increases gradually, the resonant frequencies of effective permeability shift to lower frequency bands, and the spacing between successive resonant peaks narrows simultaneously. The transmission spectrum exhibit slight spectral broadening, which is far weaker than the broadening observed in Fig. 2(a). The overall spectral response is dominated by resonant frequency drift with only mild linewidth expansion. The trends of the magnetic field intensity ratio $q_n$ agree well with those presented in Fig. 2(a). Under these parameter settings, the magnetic field profiles inside the dielectric dopant closely resemble the field distribution for the topological charge $g$=0 case and approximately follow the $\mathrm{TM}_{\mathrm{m,n}}$ eigenmodes of rectangular resonant cavities. Accordingly, we can qualitatively analyze the evolution of effective permeability by drawing on the theoretical model derived for zero topological charge. The effective permeability formula for square dielectric dopants at topological charge $g$=0 reads [35]:

$$\mu_{\mathrm{eff}} = 1 + \sum_{m=1,n=1}^{+\infty} \frac{2 l_{\mathrm{die}}^{2} \left( (-1)^{m} - 1 \right)^{2} \left( (-1)^{n} - 1 \right)^{2}}{l^{2} \pi^{4} m^{2} n^{2}} \cdot \frac{\varepsilon_{\mathrm{d}} \left( \frac{\omega}{c} \right)^{2}}{\left( \frac{m\pi}{l_{\mathrm{die}}} \right)^{2} + \left( \frac{n\pi}{l_{\mathrm{die}}} \right)^{2} - \varepsilon_{\mathrm{d}} \left( \frac{\omega}{c} \right)^{2}}, \tag{8}$$

resonant frequencies for curves describing effective permeability versus angular frequency are given by:

$$\omega_{m,n} = \frac{c}{\sqrt{\varepsilon_{\text{d}}}} \sqrt{(m\pi / l_{\text{die}})^2 + (n\pi / l_{\text{die}})^2}. \tag{9}$$

In Eqs. (8) and (9), $m$ and $n$ label the $TM_{m,n}$ eigenmodes that characterize the magnetic field profiles within the square dielectric dopant resonant cavities. From Eq. (9), for fixed mode indices $m$ and $n$, the resonant frequency $\omega_{m,n}$ decreases monotonically as the dielectric dopant side length $l_{\text{die}}$ increases, accompanied by a reduction in the frequency separation between adjacent resonant peaks. This theoretical prediction is consistent with the spectral trend observed in Fig. 2(b).The physical origins of transmission spectrum broadening differ substantially between Fig. 2(a) and Fig. 2(b). The top panels of Fig. 2(b) present normalized $z$-component magnetic field contour plots captured at each transmittance peak for different dielectric dopant sizes. When the total doped region area $A_{\text{dopant}}$ expands, the area occupied by the surrounding ENZ host medium $A_{\text{ENZ}}$ is correspondingly reduced. Nevertheless, the limited area of internal ENZ holes can only counteract a small portion of the total magnetic flux, meaning the dielectric dopant still undertakes the majority of flux cancellation. Accordingly, only a slight shift in the zero-crossing points of effective permeability away from the resonant frequencies is required to satisfy global flux balance, which yields comparatively mild spectral broadening of transmittance curves.

In summary, for photonic doping structures with topological charge $g$=1 under variable total doped region area $A_{\text{dopant}}$, two distinct tuning strategies show clear differences and complementary behaviors: fixing the dielectric dopant area $A_{\text{die}}$ while increasing the internal ENZ hole area $A_{\text{hole}}$, and fixing ENZ hole area $A_{\text{hole}}$ while expanding dielectric dopant area $A_{\text{die}}$. The first strategy induces prominent spectral broadening of the transmission spectrum, accompanied by a two-stage migration of resonant transmittance peaks: the peaks initially shift toward lower frequencies and then move to higher frequencies, where a distinct critical area $A_{\text{c}}$ separates these two distinct modulation regimes. The critical area $A_{\text{c}}$ is positively correlated with dielectric dopant area $A_{\text{die}}$ and increases as dielectric dopant area $A_{\text{die}}$ grows, while it has no correlation with the relative permittivity $\varepsilon_{\text{d}}$ of the dielectric dopant. The second strategy is dominated by low-frequency shifts of the effective permeability resonances and only induces mild transmission spectrum broadening, enabling finer regulation of resonant frequency displacement. Both strategies tailor the effective permeability and transmission spectrum by reconstructing internal magnetic field profiles and balancing mutually opposing magnetic flux components.

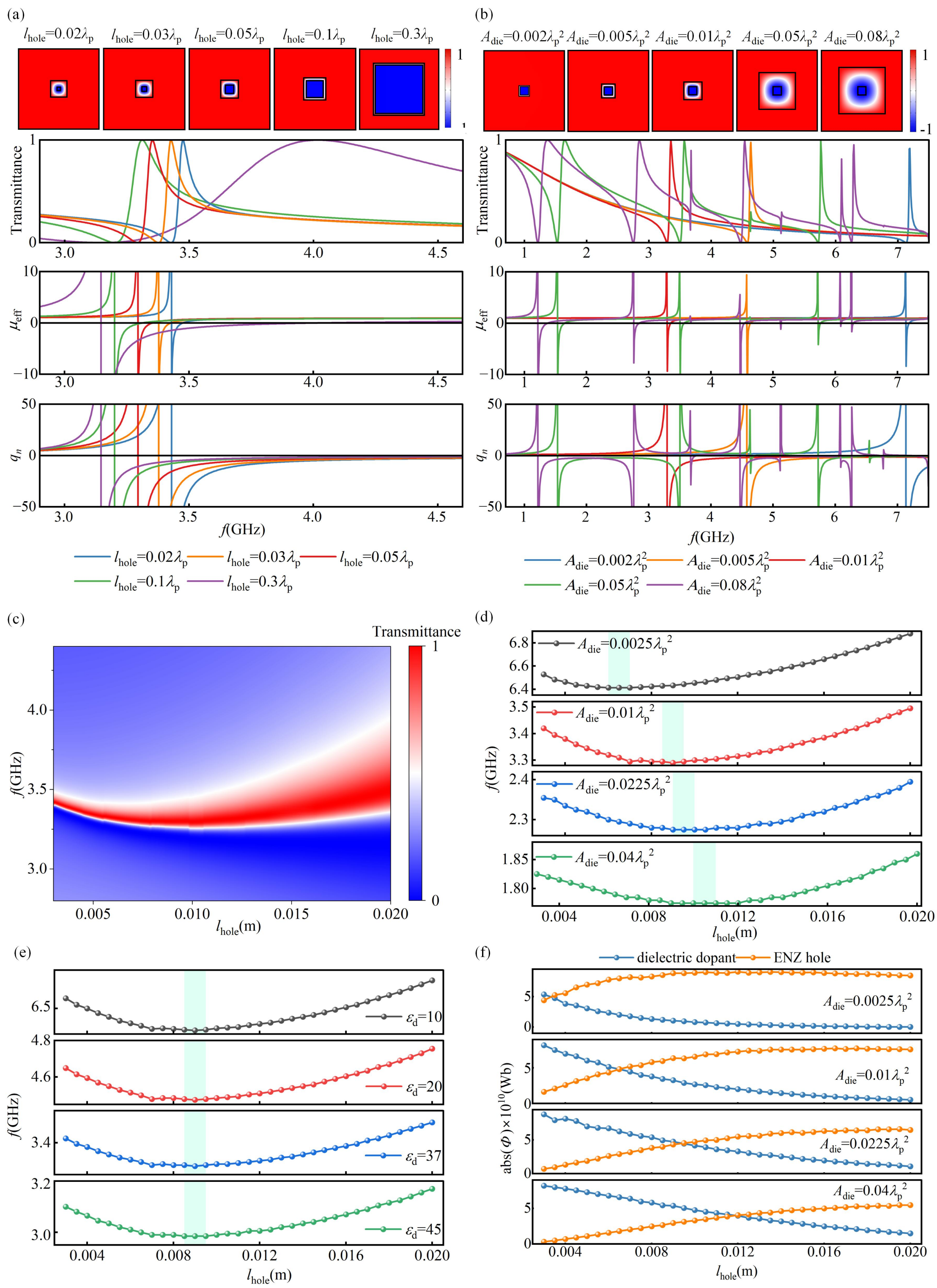

**Fig. 2.** Tuning laws governing transmission spectrum characteristics of photonic doping waveguides with variable total doped region area $A_{\mathrm{dopant}}$, as regulated by the areas of internal ENZ holes and dielectric dopants. (a) Variations in transmittance, effective permeability $\mu_{\mathrm{eff}}$, and magnetic field intensity ratio $q_n$ when adjusting the ENZ hole area $A_{\mathrm{hole}}=l_{\mathrm{hole}}\times l_{\mathrm{hole}}$ with constant dielectric dopant area $A_{\mathrm{die}}=0.01\lambda_{\mathrm{p}}^2$ and (b) adjusting the dielectric dopant area $A_{\mathrm{die}}$ with constant ENZ hole area $A_{\mathrm{hole}}=0.0025\lambda_{\mathrm{p}}^2$. Each top inset shows the normalized $z$-component magnetic field distributions within doped ENZ body evaluated at each transmittance peak under the corresponding geometric

parameters. (c) Two-dimensional color contour map illustrating how transmittance evolves with the ENZ hole side length $l_{hole}$ and incident frequency $f$. (d) Curves of transmittance peak frequency versus the ENZ hole side length $l_{hole}$ obtained on the basis of the modulation law revealed in Fig. 2(a), under different dielectric dopant areas $A_{die}$ and (e) different dielectric dopant relative permittivity $\varepsilon_d$. (f) Curves (corresponding to Fig. 2(f)) depicting the total magnetic flux of dielectric dopants and internal ENZ holes at transmittance peaks as functions of ENZ hole side length $l_{hole}$ for various dielectric dopant area $A_{die}$.

### 3.2 Tuning Laws of Hole Parameters with Fixed Total Doped Region Area

With the total doped region area $A_{dopant}$ kept constant, Figs. 3(a)-3(c) sequentially investigate how the ENZ hole side length $l_{hole}$, offset distance $s$, and hole geometry modulate transmission spectrum responses. These three parameters only induce pure resonant frequency shifts of transmission spectra without introducing spectral broadening. Increasing either ENZ hole side length $l_{hole}$ or the offset distance $s$ pushes the transmission spectra to higher frequencies. Among the four hole geometries considered, triangular holes yield the largest resonant frequency shift, whereas four-lobed holes exhibit frequency shifts comparable to those of square holes. Fig. 3(d) further quantifies these trends: the transmittance peak frequency rises nonlinearly with growing ENZ hole side length $l_{hole}$ at an accelerating rate (black curve); by contrast, the peak frequency increases sharply at first and then saturates as the offset distance $s$ becomes larger (red curve).

Physically, increasing the ENZ hole side length $l_{hole}$, enlarging the offset distance $s$, and raising the geometric mismatch between the hole contour and square geometry all amplify the internal magnetic field distortion within the dielectric dopant, thereby inducing a larger resonant frequency shift of transmittance peaks. Specifically, both a larger ENZ hole side length $l_{hole}$ and a greater offset distance $s$ shift the transmittance peaks to higher frequencies. As ENZ hole side length $l_{hole}$ increases, magnetic field distortion grows continuously, and the resonant transmittance peak frequency rises nonlinearly with a progressively steeper slope. As the offset distance $s$ increases, magnetic field distortion first rises rapidly before decaying slowly; accordingly, the peak frequency increases sharply at the initial stage, and its growth rate slows once the hole approaches the boundary of the dielectric dopant. In comparison, modifying the hole geometry can drive resonant peaks to shift toward either higher or lower frequencies, with the shift direction governed by the hole's outer contour. All three modulation strategies realize pure frequency-shift tuning that only modifies resonant peak positions, which act as independent schemes for transmission spectrum regulation.

In summary, for a fixed topological charge $g$=1 and constant total doped region area $A_{dopant}$, adjusting the size, offset distance, or geometric profile of internal ENZ holes reshapes the internal magnetic field distributions within the dielectric dopant and thereby modulates the resonant behaviors of effective permeability. This modulation only produces pure resonant frequency shifting of transmission spectrum without introducing any spectral broadening. The magnitude of resonant frequency shift is positively correlated with the degree of internal magnetic field distortion: a larger hole side length $l_{hole}$ or a larger offset distance $s$ yields a more prominent high-frequency shift of transmission peaks; the weaker the geometric resemblance to square holes, the greater the frequency offset compared with the square-hole configuration. This pure frequency-shift tuning mechanism acts as a complementary scheme to the spectral broadening effect triggered by varying the hole area or dielectric dopant area, allowing independent and accurate manipulation of transmission spectrum.

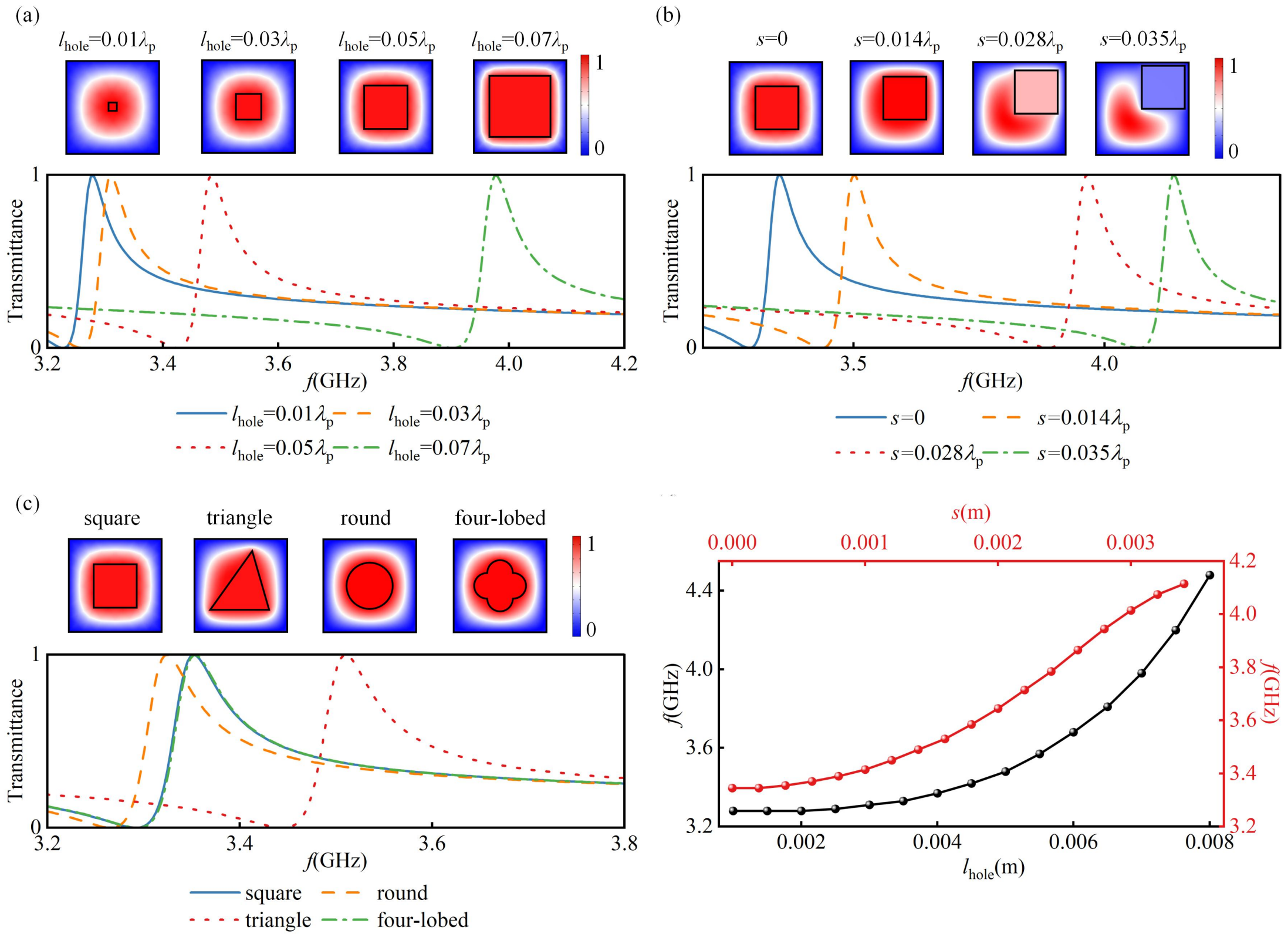

**Fig. 3.** Tuning laws governing transmission spectrum characteristics of photonic doping waveguides with a fixed total doped region area $A_{\text{dopant}}$, as modulated by the geometric parameters of internal ENZ holes. (a) Evolution of transmission spectrum when solely varying the hole side length $l_{\text{hole}}$, with all other parameters kept constant. (b) Variations in transmission spectrum obtained by tuning the offset distance $s$ (measured relative to the geometric center of dielectric dopants), and (c) internal ENZ hole geometries, where the ENZ hole area $A_{\text{hole}}$=0.0025$\lambda_p^2$. The top inset of each subplot shows the normalized $z$-component magnetic field amplitude distributions within dielectric dopants evaluated at the resonant transmission peak frequencies under the corresponding parameter sets. (d) Quantitative curves plotting resonant peak frequency as a function of ENZ hole side length $l_{\text{hole}}$ (black line) and offset distance $s$ (red line).

## 4. Conclusion

This work constructs photonic doping structures with nonzero topological charge by embedding internal ENZ holes inside dielectric dopants. Their electromagnetic responses are systematically explored via theoretical derivation and full-wave numerical simulations. Introducing topological charge $g$ offers an extra independent degree of freedom to tailor transmission spectra. The underlying physical mechanism originates from the fact that embedded ENZ holes perturb the internal magnetic field profiles within dielectric dopants, which breaks the zero-topological-charge ($g$=0) configuration corresponding to zero effective permeability. We propose two distinct categories of modulation strategies for tailoring transmission spectrum in topological-charge-enabled photonic doping systems. The first category achieves pure resonant frequency shifting without spectral broadening under a constant total doped region area $A_{\text{dopant}}$. This branch includes two sub-methods: we can adjust the topological charge by varying the number of internal ENZ holes, or alternatively reshape the internal magnetic field distributions by tuning the size, offset distance, and geometric profile of embedded holes while keeping the topological charge fixed. In both subcases, the resonant magnetic field profiles within the dielectric dopant are modified, which alters the effective permeability and further shifts the transmission peaks. The magnitude of frequency offset depends on the degree of internal magnetic field perturbation relative to the topological charge $g$=0 reference configuration: severe field distortion produces large spectral shifts, whereas weak distortion only leads to minor resonant frequency variations. The second category introduces controllable spectral broadening by varying the total doped region area $A_{\mathrm{dopant}}$. This can be realized in two complementary manners: expanding the ENZ hole area $A_{\mathrm{hole}}$ with a fixed dielectric dopant area $A_{\mathrm{die}}$, or increasing the dielectric dopant area $A_{\mathrm{die}}$ under a constant ENZ hole area $A_{\mathrm{hole}}$. Either approach changes the balance of mutually opposing magnetic fluxes inside the doped structure, thereby generating different degrees of spectral linewidth broadening. Combining these two categories of modulation schemes enables independent, simultaneous control over resonant frequency and spectral linewidth, offering unprecedented flexibility to engineer photonic doping systems with arbitrary customized transmission spectra.

## Acknowledgments

This work is supported by the National Natural Science Foundation of China (Nos. 12274317 and 12374277), San Jin Talent Support Program—Shanxi Provincial Youth Top-notch Talent Project (SJYC2025302), Natural Science Foundation of Shanxi Province (202303021211054), and Shanxi Province Higher Education Institutions Young Faculty Research and Innovation Support Program (2025Q006).

## Competing interests

The authors declare that they have no competing interests.

## Data and materials availability

All data needed to evaluate the conclusions in the paper are present in the paper and/or the Supplementary Materials.